\documentclass[prb]{revtex4}
\usepackage{graphicx,epsfig}

\begin{document}

\title{Transition from a fractional quantum Hall liquid to an electron solid at Landau level filling $\nu = 1/3$ in tilted magnetic fields}

\author{W. Pan$^{*}$, G.A. Cs\'{a}thy, and D.C. Tsui}
\affiliation{Department of Electrical Engineering, Princeton University, Princeton, New Jersey 08544}

\author{L.N. Pfeiffer and K.W. West}
\affiliation{Bell Labs, Lucent Technologies, Murray Hill, New Jersey 07954}

\vskip10pc

\begin{abstract}

We have observed in a low density two-dimensional hole system (2DHS) of extremely high quality
(with hole density $p = 1.6\times10^{10}$ cm$^{-2}$ and mobility $\mu = 0.8\times10^6$ cm$^2$/Vs) that, as the 2DHS is
continuously tilted with respect to the direction of the magnetic field, the $\nu=1/3$ fractional
quantum Hall effect (FQHE) state is weakened and its magnetoresistivity rises from $\sim 0.4$~k$\Omega/square$
in the normal orientation to $\sim 180$~k$\Omega/square$ at tilt angle $\theta \sim 80^{\circ}$. We attribute this phenomenon to the
transition of the 2DHS from the FQHE liquid state to the pinned Wigner solid state, and argue that
its origin is the strong coupling of subband Landau levels under the tilted magnetic fields.

\end{abstract}


\date{\today}
\maketitle

The phase transition from a fractional quantum Hall liquid phase \cite{tsui82} to the Wigner solid (WS) phase
\cite{lozovik75,fukuyama79} at high magnetic ($B$) fields or low Landau level fillings ($\nu$) has been a topic of much
interest in the study of the two-dimensional electron/hole system (2DES/2DHS) \cite{review}. Experimentally, these two phases
show distinctively different temperature ($T$) dependent behaviors in electronic transport. For the fractional quantum Hall
effect (FQHE) state, the magneto-resistivity ($\rho_{xx}$) becomes vanishingly small when $T$ is much lower than its energy
gap ($\Delta$), and increases with increasing temperatures. While for the WS, weakly pinned by residual impurities in real
samples, it is an insulator. Consequently, $\rho_{xx}$ decreases with increasing temperatures. Earlier theoretical studies
found that the critical transition point depends on the strength of carrier-carrier interaction (characterized by the
parameter $r_s = (\pi n)^{-1/2} m^*e^2/4\pi\hbar^2\epsilon\epsilon_0$ or $(\pi p)^{-1/2}
m^*e^2/4\pi\hbar^2\epsilon\epsilon_0$) and moves towards higher $n$ at larger $r_s$ \cite{chui91} (Here, $n$ (or $p$) is the
2DES (or 2DHS) density, and other parameters have their usual meanings). For instance, in a high quality 2DES of density $n
\sim 1\times10^{11}$ cm$^{-2}$, where $r_s \sim 2$, the WS phase becomes the ground state at the B field above that of
$\nu=1/5$ \cite{jiang90}. While in a high quality dilute 2DHS of density $p \sim 5\times 10^{10}$ cm$^{-2}$, where $r_s \sim
14$, the DC transport measurements \cite{santos92} and, especially, the microwave data \cite{li97}, in which the resonance
due to pinning of the WS was discovered at $\nu \sim 0.3$, clearly demonstrated the formation of the WS phase at higher $\nu$
in the system of larger $r_s$.

The magnitude of $r_s$ can be changed by changing the carrier density or changing the 2D system to
carriers of a different band mass ($m^*$). As shown above, with similar carrier densities, $r_s$ in the
2DHS is about 5 times larger than in the 2DES, because of a larger band mass. On the other hand,
the method of adding an in-plane magnetic field ($B_{ip}$), provided by {\it in-situ} tilting of the 2D carrier
system at low temperatures and in high magnetic fields \cite{fang68}, has not been explored in the experimental
studies, while theoretically, it was pointed out \cite{yu02} recently that a phase transition of the FQHE state
from the liquid phase to the Wigner solid phase can be induced by a non-zero $B_{ip}$. The physical origin behind
this phase transition is the effective increase of $r_s$ through the coupling of the Landau levels (or, more accurately,
magnetic levels, since the Landau level index is no longer a good quantum number with a non-zero $B_{ip}$) and 2D subbands
(or, the electric levels) \cite{yoshioka86,zhang86,yang90}. This Landau level mixing
effect reduces the difference between the energies of a
FQHE liquid state and the WS state and makes the 2D system effectively more dilute. As a result, the $r_s$ value becomes
effectively larger, and a phase transition from a FQHE state to WS can occur. So far, such phase transition has not
been identified in experiments.

In this paper, we report the observation of a crossing from the $\nu=1/3$ FQHE state to an insulating phase in a 2DHS.
When tilting the sample with respect to the direction of $B$ field, the $\nu=1/3$ FQHE state is weakened and at $T \sim 30$~mK
its magneto-resistivity ($\rho_{xx}$) rises from $\sim 0.4$~k$\Omega/square$ in the normal
orientation to $\sim 180$~k$\Omega/square$ at the tilt angle $\theta \sim 80^{\circ}$.
This value is $\sim  7 \times h/e^2$, indicating that the $\nu=1/3$ state is deeply in an insulating phase. We attribute this crossing
to the sought-after phase transition from a FQHE state to a weakly pinned Wigner solid state under tilt, and argue that
it is caused by the increased $r_s$ value through the coupling of Landau and subband levels.

Our sample consists of a modulation-doped $GaAs/Al_{0.1}Ga_{0.9}As$ quantum well (QW) of 30~nm wide, with silicon
$\delta$-doped symmetrically from both sides at a setback distance of 255~nm and grown on the \{311\}A GaAs substrate. The
2DHS density ($p$) is $p \sim 1.6\times10^{10}$ cm$^{-2}$ and varies about 5\% from one cool-down to another. The
low-temperature mobility is $\mu \sim 8\times10^5$ cm$^2$/Vsec. The band mass was measured by microwave cyclotron resonance
technique and $m^* = 0.35 m_e$ \cite{pan03}. Consequently, $r_s$ in this high quality 2DHS is $\sim 23$. Ohmic contacts to
the 2DHS were made by alloying indium-zinc (In:Zn) mixture at 440~$^{\circ}$°C for 10 minutes in the forming gas. The sample
was placed inside the mixing chamber of a dilution refrigerator of base temperature $\sim$ 30~mK (or $\sim$ 60~mK when
equipped with a rotating stage). Sample was tilted in situ from 0$^{\circ}$ up to 90$^{\circ}$ with respect to the direction
of $B$. The tilt angle, $\theta$, was determined from the shift of the resistance minimum of the integer quantum Hall effect
(IQHE) states, according to 1/cos($\theta$). Transport measurements of $\rho_{xx}$ and the Hall resistivity $\rho_{xy}$ were
carried out using standard low frequency ($\sim$ 7~Hz) lock-in techniques with an excitation current of less than 3~nA.
Results are reproducible in different cool-downs. No particular effort was made to align $B_{ip}$ with crystallographic
directions and current directions.
%

Fig.1 shows a trace of $\rho_{xx}$ vs. $B$, taken in the normal orientation. Several features are worth emphasizing:
(1) The fully-developed IQHE states at $\nu=1$ and 2 and very strong FQHE states at $\nu=1/3$ and 2/3 are observed
at very small $B$ fields, $\sim$ 2.0 and 1.0~T, respectively, manifesting the high quality of 2DHS in this specimen.
We note here that the $\nu=1/3$ FQHE state has never been observed and studied at such a low 2D hole density.
(2) A $\rho_{xx}$ minimum is seen at $\nu=2/5$ and a dip around $\nu=3/5$. (3) Between the IQHE states $\nu=1$ and 2, there is
also a dip at $\nu=5/3$. (4) The divergence of $\rho_{xx}$ beyond $\nu=1/3$ indicates that the sample is in the insulating
regime of the pinned WS phase \cite{santos92,li97}.
(5) No reentrant insulating phase is observed between $\nu=2/5$ and 1/3 in our sample.

{\it in situ} tilting was performed at $T \sim 60$~mK. In Fig.2, $\rho_{xx}$ traces at seven $\theta$'s are plotted as a
function of perpendicular $B$ field, $B_{perp} = B \times$ cos($\theta$). It is clearly seen that the $\nu=1/3$ state evolves
from a well-behaved FQHE state to a strongly insulating state. In details, at small tilt angles, e.g., from 0$^{\circ}$ to
$\theta=33^{\circ}$, the variation of $\rho_{xx}$ is small and $\nu=1/3$ is still a good FQHE state (also manifested by the
Hall resistance, not shown). Upon further tilting, $\rho_{xx}$ increases considerably. At $\theta=77^{\circ}$, $\rho_{xx}
\sim 36$~k$\Omega/square$ and has exceeded $h/e^2$ ($\sim$ 26~k$\Omega/suare$), signaling that the 2DHS has entered into an
insulating phase. $\rho_{xx}$ continues to increase, to 128~k$\Omega/square$ $\sim 5 \times h/e^2$ at $\theta=80^{\circ}$.
%

In Fig.3, $\rho_{xx}$ at $\nu=1/3$ is plotted as a function of in-plane $B$ field, $B_{ip} = B \times$ sin($\theta$).
Overall, $\rho_{xx}(\nu=1/3)$ increases exponentially with $B_{ip}$, $i.e.$, $\rho_{xx}(\nu=1/3) \propto $ exp($B_{ip}/B_0)$.
At $T \sim 60$~mK, $B_0 =3.6$~T.

In Fig.4, the temperature dependence of $\rho_{xx}$ at $\nu=1/3$ is shown on a semilog scale for the seven tilt angles. At
$\theta = 0^{\circ}$, $\rho_{xx}$ decreases as $T$ decreases from $T \sim 200$~mK to $\sim$ 60~mK (or $1/T$ increases from
$\sim$ 5 to $\sim$ 16 K$^{-1}$), the characteristic of a FQHE liquid. From a fit to the linear portion of data, an energy gap
of $\Delta \sim 0.2$~K is obtained. The energy gap decreases with increasing $\theta$ and at $\theta = 61^{\circ}$,
$\rho_{xx}$ is nearly temperature independent, indicating a zero energy gap at $\nu=1/3$. When $\theta$ is further increased,
on the other hand, $\rho_{xx}$ increases as $T$ decreases, the signature of an insulating phase. In Fig.4b, a linear fit to
the higher $T$ data points at $\theta = 80^{\circ}$, using the formula $\rho_{xx} \propto$ exp($E_g/2k_BT$), yields a
characteristic energy scale of $E_g \sim 0.4$~K.
%

Fig.5 shows the temperature dependence of $\rho_{xx}$ at $\nu=2/3$. Similar to the $\nu=1/3$ state, the strength of the
$\nu=2/3$ state shows little changes at small tilt angles. Only when $\theta \geq 71^{\circ}$, it becomes weaker, probably
the precursor of the same crossing from a FQHE state to an insulator. In contrast, the $\nu=2/5$ and 3/5 states disappear
quickly at small tilts, becoming indiscernible at $\theta=45^{\circ}$. As for the IQHE states, no weakening of the $\nu=1$
and 2 states is observed over the whole tilt range. Finally, it is interesting to note that for the peak at $B_{perp} \sim
1.8$~T and $\nu \sim 0.36$, its resistance increases continuously as $\theta$ increases, from $\sim$~8 k$\Omega/square$ at
$\theta = 0^{\circ}$ to $\sim 230$~k$\Omega/square$ at $\theta = 80^{\circ}$.

The data in Fig.2 and Fig.4 clearly demonstrate that, as $\theta$ is increased, the $\nu=1/3$ state evolves from a FQHE
liquid in the normal $B$ field orientation to a hole insulator at high tilt angles. Considering the nature of the strong
correlation in its precursor, the $\nu=1/3$ FQHE state, and its proximity to the WS phase at yet higher $B$ fields, we assign
this hole insulator to the pinned WS phase. This is for the first time that the robust $\nu=1/3$ FQHE state is destroyed by
the tilted magnetic field and becomes a WS phase at high tilt angles. The lack of such observation in previous experiments
indicates that a high quality dilute 2DHS and consequently a large $r_s$ are important for this observation. As to the
mechanism for the formation of this tilt induced WS phase, we believe that it is the effective increase of $r_s$ through the
coupling of Landau levels and 2D hole subbands under tilt. It is well known that for an ideal 2D system a tilted magnetic
field does not modify the orbital motion but only the Zeeman splitting. A real 2D electron system has finite thickness of ~10
nm, and the orbital motion is affected only to the second order: The in-plane magnetic field squeezes the electron
wavefunction and makes the electron system more two-dimensional. Indeed, the energy gap of the $\nu=1/3$ FQHE state was found
only to increase slightly \cite{boebinger87} with increasing tilt angles. In contrast, for a 2DHS, because of the
non-parabolic nature of its valence band and the spin-orbital interaction, in the presence of $B_{ip}$, the orbital motion is
affected greatly and there exists a strong coupling of Landau levels and 2D subbands \cite{maan84,
brummel86,heuring89,halonen90,goldoni93}. As pointed out in earlier studies \cite{yoshioka86,santos92}, Landau level mixing
makes the 2D system effectively more dilute and therefore, $r_s$ becomes effectively larger. Consequently, the difference
between the energies of a FQHE liquid state and the WS state is reduced \cite{yu02,yoshioka86,zhang86,yang90}. In addition to
Landau level mixing, an increment of effective mass $m^*$ under in-plane $B_{ip}$ \cite{smrcka94,aikawa02} also directly
contributes to increasing $r_s$, which is proportional to $m^*$. Following the previously proposed phase diagram of $\nu$ vs.
$r_s$ \cite{chui91}, a crossing from the FQHE state to the WS phase is then possible at $\nu=1/3$ by increasing $r_s$ (or the
tilt angle), as shown by the arrow in the inset of Fig.2. This evolution can also be viewed as that with increasing $r_s$ the
onset of the WS phase moves towards higher $\nu$. Indeed, in our measurements, the critical transition point ($\nu_c$) from
the 2D hole liquid phase to the WS phase, identified as the temperature-independent point in the traces of $\rho_{xx}$ vs.
$T$, moved from $\nu_c \sim 0.32$ at $\theta = 0^{\circ}$ to higher $\nu$ as the tilt angle was increased and, at $\theta =
80^{\circ}$, $\nu_c \sim 0.62$, just above the field of the $\nu=2/3$ state. Finally, we speculate that the similar mechanism
may also be responsible for the transition at the resistance peak of $B_{perp} \sim 1.8$~T, which was in a metallic phase
($\rho_{xx} \sim 8$~k$\Omega/square$) at the zero tilt and became an insulator ($\rho_{xx} \sim 230$ k$\Omega/square$) at
large tilt angles.

Next, we notice that even deep in the insulating regime, $\rho_{xx}$ still shows a local minimum at $\nu=1/3$, riding on a
huge background. The origin of this local minimum may be an indication of the coexistence of the FQHE liquid and the WS. It
is known that even in this highest quality 2DHS, residual impurities and density fluctuations are inevitable. As a result,
the 2DHS may break into domains of liquid and WS. Percolation of the FQHE liquid through the pinned WS can give rise to a
local resistance minimum. On the other hand, this resistance dip may be due to electron-electron interactions. It has been
shown that the inclusion of the Laughlin-Jastrow correlations in the Wigner crystal regime can lower the ground state energy
near a rational Landau filling \cite{yi98}.  This energy lowering may also be responsible for the local resistance minimum at
$\nu=1/3$.

Before we finish the paper, we want to point out that, firstly, the formation of the insulting phase is unlikely caused by
the enhanced disorder scattering seen by the 2D holes. It is known that under a non-zero in-plane magnetic field the
effective mass ($m^*$) of the 2D carrier increases and the effective scattering time $\tau$ decreases. Consequently, the
effective mobility $\mu=e\tau/m^*$ decreases and the resistivity increases. The increase of $m^*$ is believed to be due to
the coupling of in-plane $B$ field to the carrier orbital motion \cite{smrcka94}. Recently in a two-dimensional electron
system, a 20\% increase of $m^*$ was observed \cite{aikawa02}. The decrease of $\tau$ is due to the enhanced surface
roughness scattering, since under non-zero $B_{ip}$ the 2D hole wavefunction is squeezed and pushed closer to the interface
of $GaAs/AlGaAs$. We have shown that the coupling effect is responsible for the observed insulating phase. The enhanced
roughness scattering mechanism cannot be the origin of the high tilt insulating phase. In detail, at $\theta = 90^{\circ}$,
where the 2DHS is expected to experience the maximum interface roughness scattering, a relatively small increase in
$\rho_{xx}$, from $\sim0.5$~k$\Omega/square$ at $B=0$~T to $\sim 2$~k$\Omega/square$ at $B=10$~T, was measured (as shown in
Fig.6). This factor of 4 increase is similar to that at the same density in an earlier theoretical calculation
\cite{dassarma99}, where only the coupling effect was taken into account. Furthermore, the factor of 4 increase is much
smaller than the observed $\rho_{xx}$ increase at $\nu=1/3$ from $\theta=0^{\circ}$ to $\theta=80^{\circ}$, which is more
than two orders of magnitude. Taken these results together, it can be concluded that the enhanced surface roughness
scattering is negligible.

Secondly, the increased Zeeman energy in the tilted magnetic field cannot explain the formation of the
insulating phase, either.  The non-zero in-plane magnetic field increases the total magnetic field seen
by the spin, and therefore increases the Zeeman energy. Such a variation in $E_Z$ has been used successfully
to interpret the angular dependent disappearance and reappearance of energy gaps of the spin-unpolarized
FQHE states, e.g., at $\nu=2/3$ \cite{haug87,syphers88,clark89,eisenstein89,davies91,engel92,du95,kang97,muraki99,smet01}.
However, the ground state of the $\nu=1/3$ FQHE state is known to be
fully spin-polarized. Thus, all the spins have aligned along the external $B$ field direction already at
zero tilt, and the $\nu=1/3$ state should not be affected and de-stabilized by increasing the Zeeman energy.
So far, studies in the 2DES have not identified any spin-related phase transitions at $\nu=$1/3. Furthermore,
the lack of phase transition in $\rho_{xx}$ at $\nu=2/3$ also points out that the formation of the insulating phase
is not a spin effect.

To summarize, in this paper, we report experimental results in a 2DHS of density $p = 1.6\times10^{10}$ cm$^{-2}$ and
mobility $\mu = 0.8\times10^6$ cm$^2$/Vs in the tilted magnetic fields: The $\nu=1/3$ FQHE state was weakened and
its magnetoresistivity rose from $\sim 0.4$~k$\Omega/square$ in the normal orientation to $\sim 180$~k$\Omega/square$
at the tilt angle $\theta \sim 80^{\circ}$. We attribute this crossing to a phase transition
from a FQHE liquid phase to a pinned Wigner
solid phase, and argue that this phase transition is due to the effective increase of $r_s$ value through
the strong coupling of subband Landau levels in the presence of non-zero in-plane magnetic fields.

We would like to thank H. Noh, E. Palm, and T. Murphy for technical help, and one of our referees for pointing out the
Laughlin-Jastrow correlations as an alternative explanation to the resistivity minimum at $\nu=1/3$ in the Wigner crystal
regime. A portion of this work was performed at the NHMFL, which is supported by NSF Cooperative Agreement No. DMR-9527035
and by the State of Florida. The work at Princeton was supported by the AFOSR, the DOE, and the NSF.

\vskip1pc
$^*$ Present address: Sandia National Laboratories; email: wpan@sandia.gov.

\begin{figure} [h]
\centerline{\epsfig{file=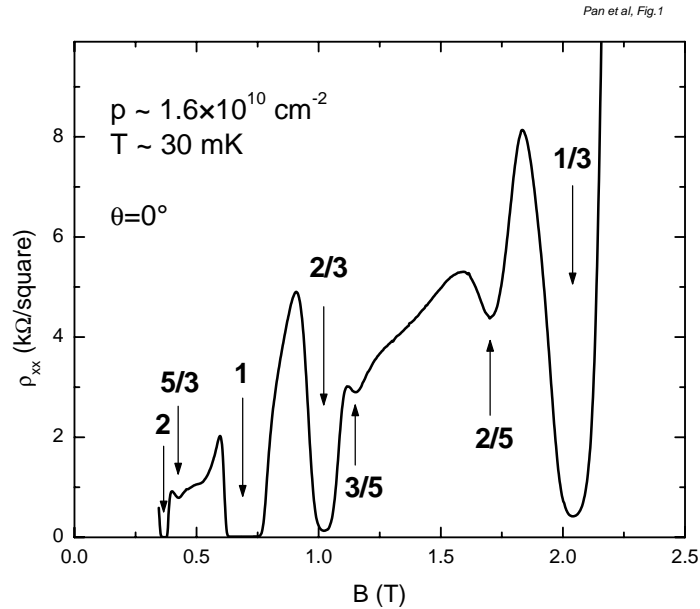,width=12.5cm}} \caption{ Overview of $\rho_{xx}$ at $\theta=0^{\circ}$. The 2DHS density is
$p \sim 1.6\times10^{10}$ cm$^{-2}$ and $T \sim 30$~mK. Major IQHE and FQHE states are indicated by the arrows.}
\end{figure}

\begin{figure} [h]
\centerline{\epsfig{file=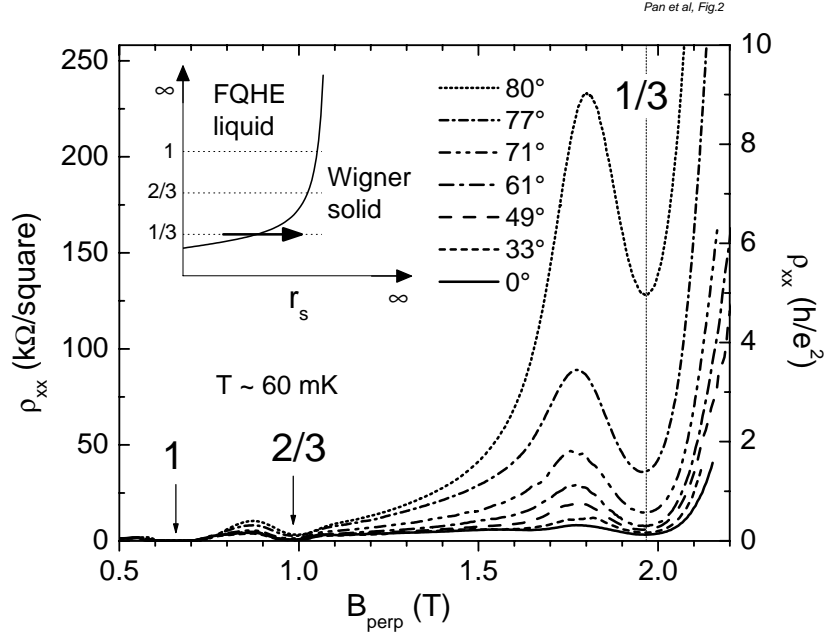,width=12.5cm}} \caption{ Tilted $B$ field dependence of $\rho_{xx}$ at seven angles at
$\theta = 0^{\circ}$, 33$^{\circ}$, 49$^{\circ}$, 61$^{\circ}$, 71$^{\circ}$, 77$^{\circ}$, and 80$^{\circ}$ (from bottom to
top). All the traces were taken at $T \sim 60$~mK, except the one at $\theta = 80^{\circ}$, which was taken at $T \sim
47$~mK. The x-axis is the perpendicular $B$ field, $B_{perp}=B\times$ cos($\theta$). The right y-axis is in units of $h/e^2$,
for the same traces. The inset shows schematically the evolution from a FQHE liquid phase (e.g., at $\nu=1/3$) to the WS
phase as $r_s$ is increased. This phase diagram is adopted from Ref. [5].}
\end{figure}

\begin{figure} [h]
\centerline{\epsfig{file=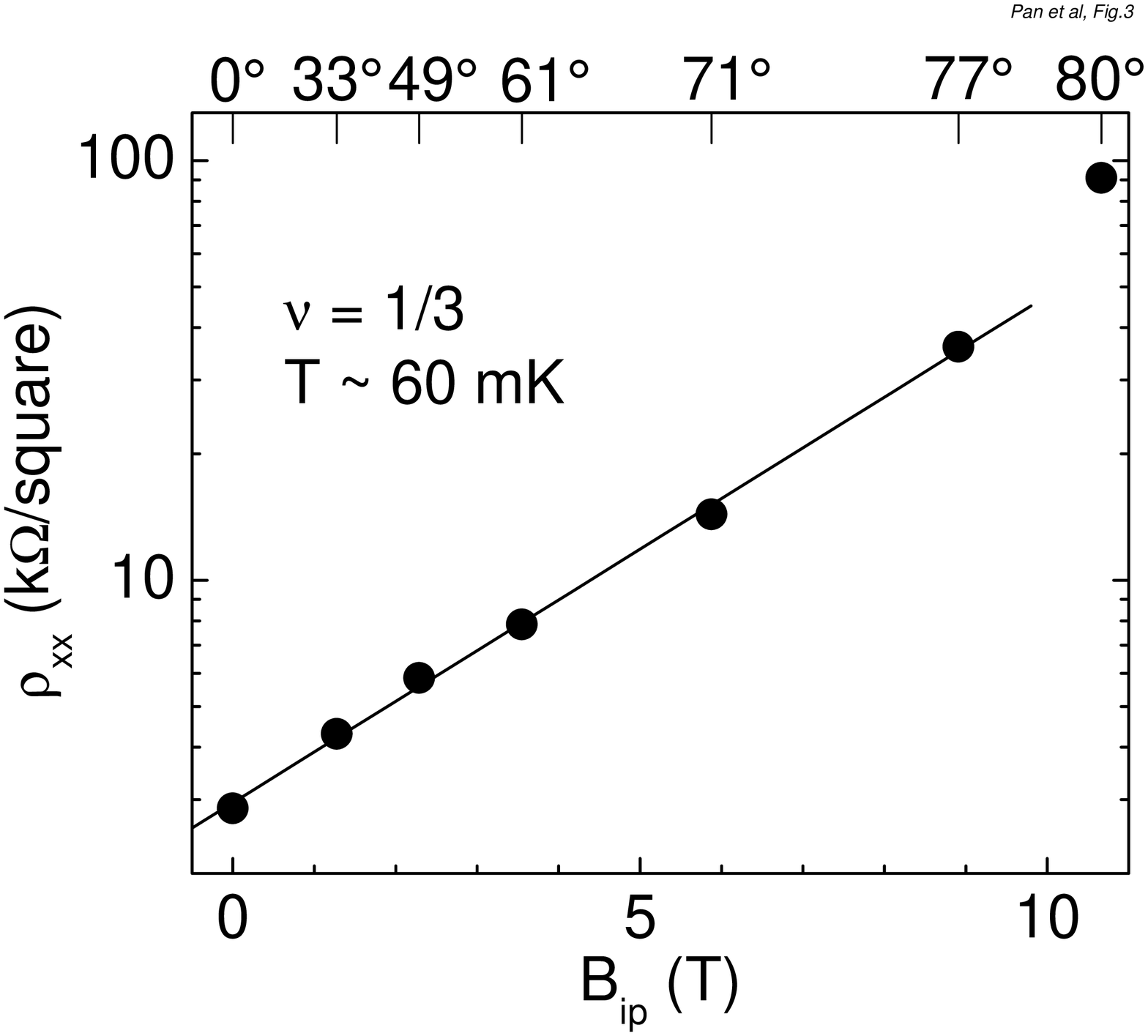,width=8.5cm}} \caption{$\rho_{xx}$ minimum at $\nu=1/3$ vs. $B_{ip}$, plotted on a semi-log
scale. The straight line is a linear fit to all data points, except the one at $\theta=80^{\circ}$.}
\end{figure}

\begin{figure} [h]
\centerline{\epsfig{file=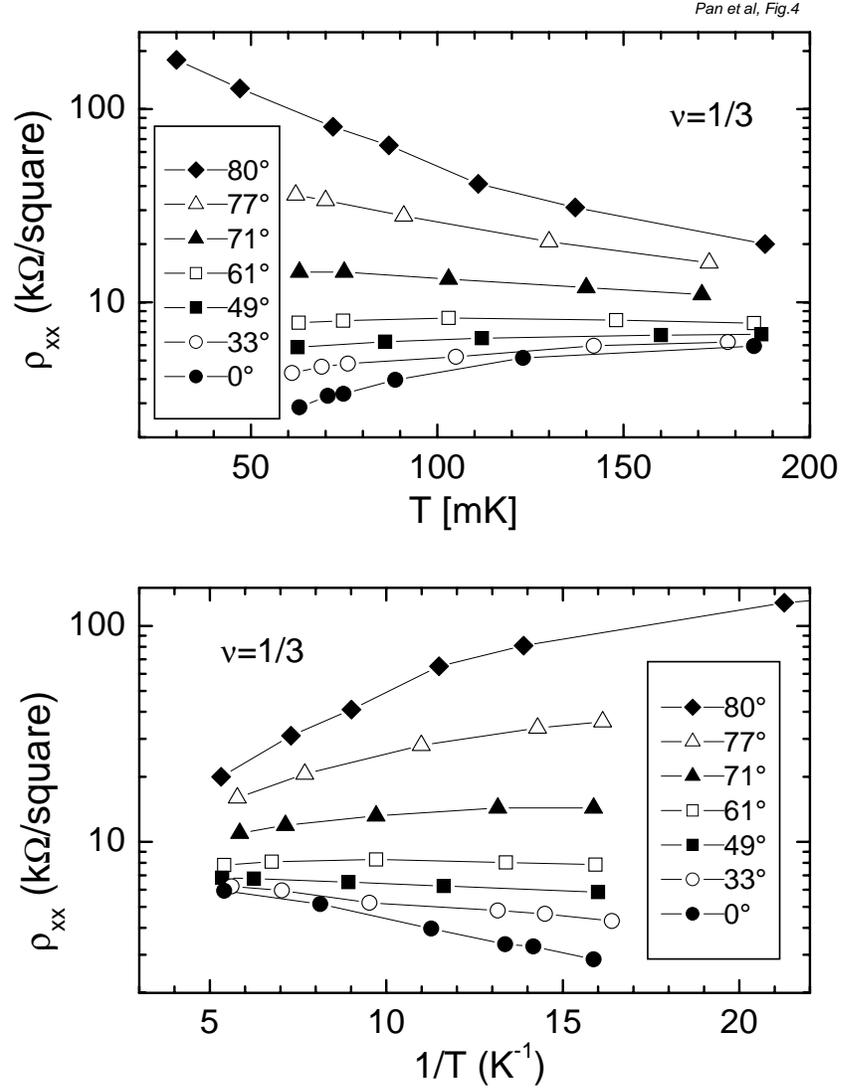,width=12.5cm}} \caption{ Temperature dependence of $\rho_{xx}$ minimum at $\nu=1/3$ at
different tilt angles: (a) $\rho_{xx}$ vs. $T$; (b) $\rho_{xx}$ vs. $1/T$ for the same data.}
\end{figure}

\begin{figure} [h]
\centerline{\epsfig{file=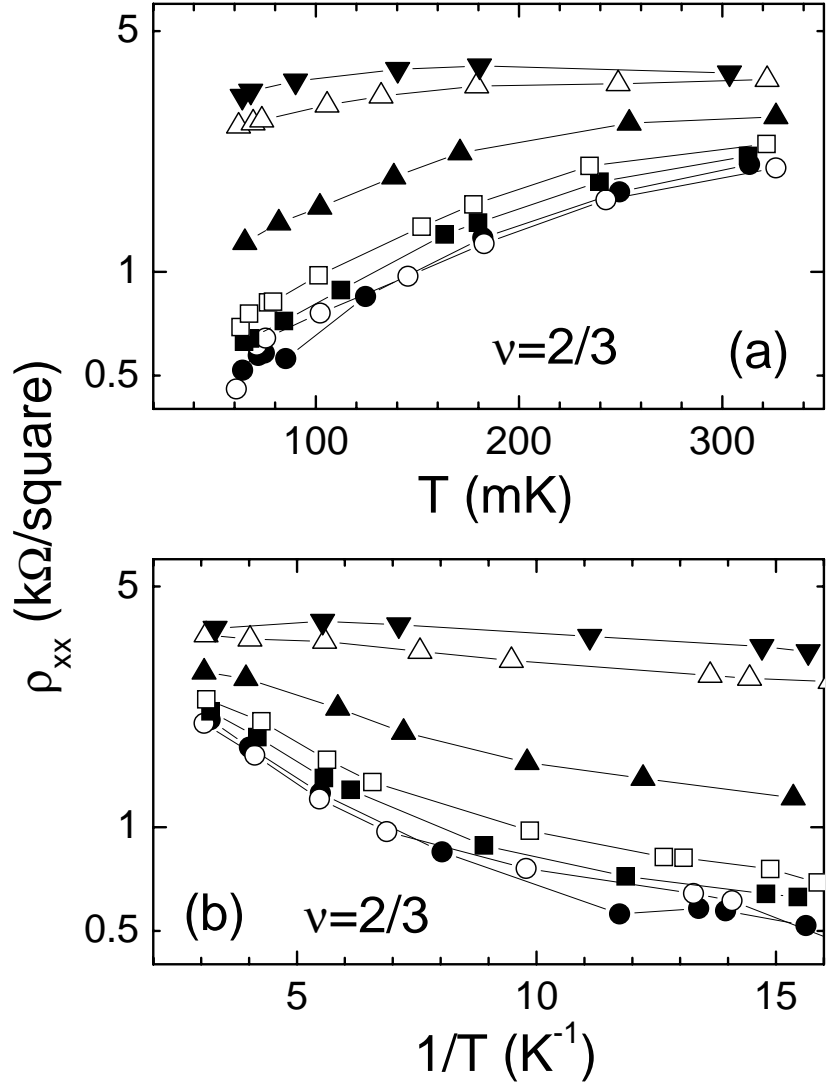,width=12.5cm}} \caption{Temperature dependence of $\rho_{xx}$ minimum at $\nu=2/3$ at
different tilt angles: (a) $\rho_{xx}$ vs. $T$; (b) $\rho_{xx}$ vs. $1/T$. The same symbol represents the same tilt angle as
in Fig.4a and Fig.4b.}
\end{figure}

\begin{figure} [h]
\centerline{\epsfig{file=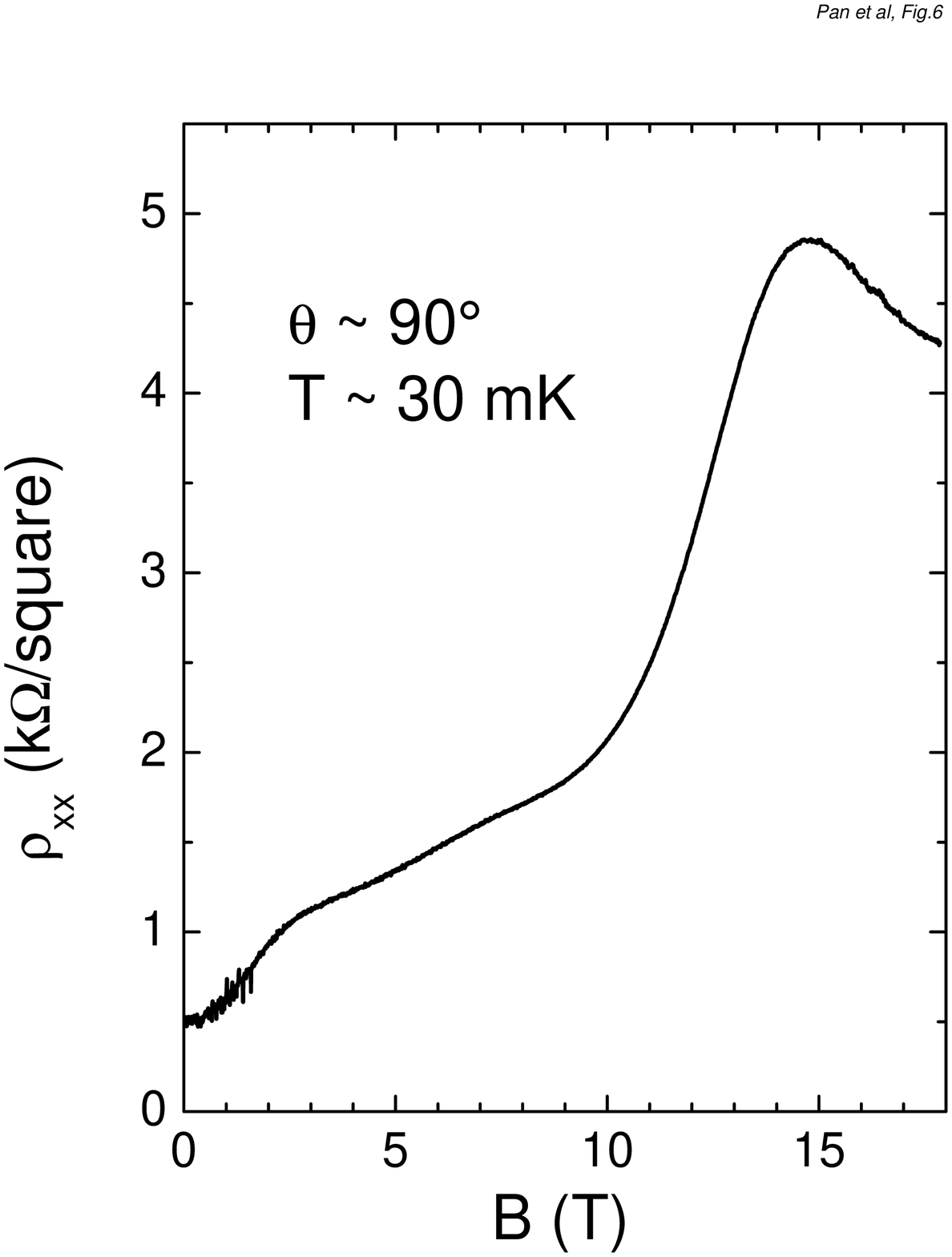,width=12.5cm}} \caption{$\rho_{xx}$ in parallel $B$ field, $\theta \sim 90^{\circ}$.}
\end{figure}


\begin{thebibliography} {90}

\bibitem{tsui82}
D.C. Tsui, H.L. Stormer, and A.C. Gossard, Phys. Rev. Lett. {\bf 48}, 1559 (1982).

\bibitem{lozovik75}
Y.E. Lozovik and V.I. Yudson, JEPT Lett. {\bf 22}, 11 (1975).

\bibitem{fukuyama79}
H. Fukuyama and P.A. Lee, Phys. Rev. B {\bf 18}, 6245 (1979).

\bibitem{review}
For a review of recent theoretical and experimental results on Wigner crystal, see,
for example, the chapters by H.A. Fertig and M. Shayegan in Perspectives in Quantum Hall Effect,
S. Das Sarma and A. Pinczuk (Eds.), Wiley, New York (1996), and references therein.

\bibitem{chui91}
S.T. Chui and K. Esfarjani, Europhys. Lett. {\bf 14}, 361 (1991).

\bibitem{jiang90}
H.W. Jiang, R.L. Willett, H.L. Stormer, D.C. Tsui, L.N. Pfeiffer, and K.W. West, Phys. Rev. Lett. {\bf 65}, 633 (1990).

\bibitem{santos92}
M.B. Santos, Y.W. Suen, M. Shayegan, Y.P. Li, L.W. Engel, and D.C. Tsui, Phys. Rev. Lett. {\bf 68}, 1188 (1992).

\bibitem{li97}
C.-C. Li, L. W. Engel, D. Shahar, D. C. Tsui, and M. Shayegan, Phys. Rev. Lett. {\bf 79}, 1353 (1997).

\bibitem{fang68}
F.F. Fang and P.J. Stiles, Phys. Rev. {\bf 174}, 823 (1968).

\bibitem{yu02}
Yue Yu and Shijie Yang, Phys. Rev. B {\bf 66}, 245318 (2002).

\bibitem{yoshioka86}
D.Yoshioka, J. Phys. Soc. Jpn. {\bf 53}, 3740 (1984); ibid. J. Phys. Soc. Jpn. {\bf 55}, 885 (1986).

\bibitem{zhang86}
F.C. Zhang and S. Das Sarma, Phys. Rev. B {\bf 33}, 2903 (1986).

\bibitem{yang90}
S.-R. Eric Yang, A.H. MacDonald, and D. Yoshioka, Phys. Rev. B {\bf 41}, 1290 (1990).

\bibitem{pan03}
W. Pan, K. Lai, S. Baraci, N.P. Ong, D.C. Tsui, L.N. Pfeiffer, and K.W. West, Appl. Phys. Lett. (2003).

\bibitem{boebinger87}
G.S. Boebinger, H.L. Stormer, D.C. Tsui, A.M. Chang, J.C.M. Hwang, A.Y. Cho, C.W. Tu, and G. Weimann, Phys. Rev. B {\bf 36},
7919 (1987).

\bibitem{maan84}
J.C. Maan, in Two Dimensional Systems, Heterostructures, and Superlattices,
edited by G. Bauer, F. Kuchar, and H. Heinrich (Springer-Verlag, Berlin, 1984).

\bibitem{brummel86}
M.A. Brummel, M.A. Hopkins, R.J. Nicholas, J.C. Portal, K.Y. Cheng, and A.Y. Cho, J. Phys. C: Solid State Phys. {\bf 19},
L107 (1986).

\bibitem{heuring89}
W. Heuring, E. Bangert, G. Landwehr, G. Weimann, and W. Schlapp,
in High Magnetic Fields in Semiconductor Physics II, edited by G. Landwehr (Springer-Verlag, Berlin, 1989).

\bibitem{halonen90}
V. Halonen, P. Pietilainen, and T. Chakraborty, Phys. Rev. B {\bf 41}, 10202 (1990).

\bibitem{goldoni93}
G. Goldoni and A. Fasolino, Phys. Rev. B {\bf 48}, 4948 (1993).

\bibitem{smrcka94}
L. Smr\u{c}ka and T. Jungwirth, J. Phys.: Condens. Matter {\bf 6}, 55 (1994).

\bibitem{aikawa02}
H. Aikawa, S. Takaoka, K. Oto, K. Murase, T. Saku, Y. Hirayama, S. Shimomura, and S. Hiyamizu, Physica E {\bf 12}, 578(2002).

\bibitem{yi98}
Hangmo Yi and H.A. Fertig, Phys. Rev. B {\bf 58}, 4019 (1998).

\bibitem{dassarma99}
S. Das Sarma and E.H. Hwang, Phys. Rev. Lett. {\bf 84}, 5596 (2000).

\bibitem{haug87}
R. J. Haug, K. v. Klitzing, R. J. Nicholas, J. C. Maan, and G. Weimann, Phys. Rev. B {\bf 36}, 4528 (1987).

\bibitem{syphers88}
D. A. Syphers and J. E. Furneaux, Solid State Commun. {\bf 65}, 1513 (1988).

\bibitem{clark89}
R.G. Clark, S.R. Haynes, A.M. Suckling, J.R. Mallett, P.A. Wright, J.J. Harris, C.T. Foxon, Phys. Rev. Lett. {\bf 62}, 1536
(1989).

\bibitem{eisenstein89}
J.P. Eisenstein, H.L. Stormer, L.N. Pfeiffer and K.W. West, Phys. Rev. Lett. 62, 1540 (1989); ibid, Phys. Rev. B {\bf 41},
7910 (1990).

\bibitem{davies91}
A.G. Davies, R. Newbury, M. Pepper, J.E.F. Frost, D.A. Ritchie, and G.A.C. Jones, Phys. Rev. B {\bf 44}, 13128 (1991).

\bibitem{engel92}
L.W. Engel, S.W. Hwang, T. Sajoto, D.C. Tsui and M. Shayegan, Phys. Rev. B {\bf 45}, 3418 (1992).

\bibitem{du95}
R.R. Du, A.S. Yeh, H.L. Stormer, D.C. Tsui, L.N. Pfeiffer, and K.W. West, Phys. Rev. Lett. {\bf 75}, 3926 (1995).

\bibitem{kang97}
W. Kang, J. B. Young, S. T. Hannahs, E. Palm, K. L. Campman, and A. C. Gossard, Phys. Rev. B {\bf 56}, 12776 (1997).

\bibitem{muraki99}
K. Muraki and Y. Hirayama, Phys. Rev. B {\bf 59}, 2502 (1999).

\bibitem{smet01}
J.H. Smet, R. Deutschmann, W. Wegscheider, G. Abstreiter, and K. von Klitzing, Phys. Rev. Lett. {\bf 86}, 2412 (2001).

\end{thebibliography}
\end{document}